\documentclass[12pt]{article}
\usepackage{epsf}
\usepackage{amsmath}
\usepackage[dvipdfmx]{graphicx}
\usepackage{color}
\usepackage{cite}
\usepackage{bm}
\usepackage{url}
\usepackage[colorlinks,citecolor=blue]{hyperref}
\usepackage{aas_macros}
\renewcommand{\thefootnote}{\#\arabic{footnote}}

\newcommand{\lesssim}{ \mathop{}_{\textstyle \sim}^{\textstyle <} }

\renewcommand{\thefootnote}{\fnsymbol{footnote}}
\def\thefootnote{\fnsymbol{footnote}}

\begin{document}

\begin{titlepage}

\begin{center}

\vskip .75in

{\Large \bf $H_0$ tension without CMB data: Beyond the $\Lambda$CDM }

\vskip .75in

{\large
Fumiya Okamatsu$\,^{1}$,  Toyokazu Sekiguchi$\,^{2}$ and  Tomo Takahashi$\,^{3}$
}

\vskip 0.25in

{\em
$^{1}$Graduate School of Science and Engineering, Saga University, Saga 840-8502, Japan
\vspace{2mm} \\
$^{2}$ Theory Center, IPNS, KEK, Tsukuba 305-0801, Japan
\vspace{2mm} \\
$^{3}$Department of Physics, Saga University, Saga 840-8502, Japan
}

\end{center}
\vskip .5in

\begin{abstract}

We investigate the $H_0$ tension in a range of extended model frameworks beyond the standard $\Lambda$CDM  without the data from the cosmic microwave background (CMB). Specifically, we adopt the data from baryon acoustic oscillations, big bang nucleosynthesis, and type Ia supernovae as indirect measurements of $H_0$ to study the tension. We show that the estimated value of $H_0$ from indirect measurements is overall lower than that from direct local ones regardless of the data sets and a range of extended models to be analyzed, which indicates that, although the significance of the tension varies depending on models,  the $H_0$ tension persists in a broad framework beyond the standard $\Lambda$CDM model even without CMB data.

\end{abstract}

\end{titlepage}

\renewcommand{\thepage}{\arabic{page}}
\setcounter{page}{1}
\renewcommand{\thefootnote}{\#\arabic{footnote}}
\setcounter{footnote}{0}

\section{Introduction \label{sec:intro}}

The Hubble constant $H_0$ is one of the most important cosmological parameters and has been measured by various ways. One such observations is cosmic microwave background (CMB) which indirectly measures it with high precision the data from Planck satellite derived   $H_0 = (67.4 \pm 0.5) $~km/sec/Mpc \cite{Aghanim:2018eyx} and the combination of the data from ACT and WMAP gives $H_0 = (67.7 \pm 1.1) $~km/sec/Mpc \cite{Aiola:2020azj}. One can also measure $H_0$ directly from local measurements such as the ones based on a distance ladder and strong gravitational lensing observations, which actually  gives a value of $H_0$  higher than that obtained from CMB. For example, the Cepheid-supernovae distance ladder provides $H_0 = (73.2 \pm 1.3)$~km/sec/Mpc \cite{Riess:2020fzl} (for earlier results, see \cite{Riess:2019cxk,Riess:2018uxu,Riess:2016jrr}). Although a measurement from Tip of the Red Giant Branch obtained an intermediate value for $H_0$ somewhat between CMB and the local one \cite{Freedman:2020dne}, other observations such those using  Mira variables \cite{Huang:2019yhh}, the Tully-Fisher relation \cite{Kourkchi:2020iyz,Schombert:2020pxm},  megamaser \cite{Pesce:2020xfe},  and the gravitational lensing from H0LiCOW  \cite{Wong:2019kwg}  and STRIDES \cite{Shajib:2019toy}  have also derived similar values with the above mentioned Cepheid-supernovae distance ladder, which are discrepant with the one obtained by CMB about 5$\sigma$ \cite{Riess:2020sih} (see also \cite{Perivolaropoulos:2021jda} for the compilation of various measurements). This inconsistency is now called the $H_0$ tension, which has been a target of intense study recently. Although the tension might be attributed to some unknown systematics (for the arguments of the systematics in distance ladder, see e.g., \cite{Efstathiou:2020wxn},  and in strong gravitational lensing, see e.g., \cite{Birrer:2020tax}), it could indicate that we need to extend/modify the standard concordance model of cosmology, a flat $\Lambda$CDM model, to resolve the tension and many works have been performed along this line (see, e.g.,  \cite{Knox:2019rjx,DiValentino:2021izs,Perivolaropoulos:2021jda} for a review and references therein for models to solve the tension proposed so far).  Furthermore, it is worth mentioning that the $H_0$ tension can also make some impact on other aspects of cosmology such as cosmological bounds on neutrino mass \cite{Sekiguchi:2020igz}.

Actually, the tension in $H_0$ has been mainly addressed as the discrepancy between CMB and local direct measurements.  Although CMB is the most powerful indirect measurement of $H_0$, if the tension is a genuine one, it would be persistent even without CMB data, which should be checked against various other observations.  Indeed,  some works along this line have been done such as the ones using weak lensing data from DES and baryon acoustic oscillation (BAO) with big bang nucleosynthesis (BBN)  \cite{Abbott:2017smn},  BAO and BBN \cite{Addison:2017fdm,Cuceu:2019for,Schoneberg:2019wmt,eBOSS:2020yzd}, the full shape of galaxy power spectrum and BAO with BBN \cite{Philcox:2020vvt}, BAO, type Ia supernovae (SNeIa) and cosmic chronometers \cite{Benisty:2020otr}, all of which do not use CMB data,  but the value of $H_0$ are overall consistent with the one obtained from Planck. This indicates that the $H_0$ tension exists even without CMB. However, such analysis have been  performed mainly in the framework of the standard $\Lambda$CDM  and some extended models such as the ones with neutrino masses $\sum m_\nu$  and the effective number of neutrinos (dark radiation) $N_{\rm eff}$.  In fact,  it has been known that simple extensions of $\Lambda$CDM model such as the ones with varying dark energy  equation of state,  dark radiation, neutrino masses and so on cannot fully resolve the tension\footnote{
Although these extended models cannot fully resolve the tension, they can reduce its significance (see, e.g., the discussion in \cite{Vagnozzi:2019ezj}).
}, however, those extended models have been mainly investigated using CMB  in combination with some other data sets.   As mentioned above, it would be important to check whether the $H_0$ tension is persistent or not even without CMB data, and besides it should be studied in various extensions of $\Lambda$CDM model, which is the prime purpose of this paper. To elucidate whether the $H_0$ tension exists without CMB data in various frameworks beyond  $\Lambda$CDM, in this paper we investigate the constraint on $H_0$ from BAO, SNeIa and BBN in models with extended phenomenological  dark energy, the curvature of the Universe and the effective number of extra radiation species. If the value of $H_0$ obtained in these analyses is consistent with the one obtained from CMB and deviates from the ones obtained by the local direct measurements, the $H_0$ tension would be rigorously confirmed, which may indicate that we need  to consider an intricate extension/modification of the standard cosmological model more seriously. 

The organization of this paper is as follow. In the next section, we explain the methodology of our analysis where the analysis method and the data adopted in this paper are presented. We also summarize the model framework beyond $\Lambda$CDM that we consider in this paper. In Sec.~\ref{sec:results}, we present our results and discuss the $H_0$ tension from the data without CMB in models beyond $\Lambda$CDM. The final section is devoted to our conclusion  and discussion.

\section{Methodology \label{sec:methodology}}

Here we describe the methodology of our analysis to study the $H_0$ tension without CMB data in the frameworks beyond the $\Lambda$CDM model. First we explain our method to derive constraints on $H_0$, other cosmological parameters, and the data adopted in our analysis. Then we describe the models analyzed in this paper.

\subsection{Analysis method}

To make a parameter estimation, we perform a Markov Chain Monte Carlo (MCMC) analysis using {\tt emcee} \cite{2013PASP..125..306F}.  We adopt the data from BAO, BBN and SNeIa in our analysis.  
In the BAO analysis,  some combinations of the quantities such as the Hubble distance $D_H (z)$, the line-of-sight comoving distance $D_C (z)$, the (comoving) angular diameter distance $D_M (z)$  and the sound horizon at the drag epoch $r_d$ are measured. Here the Hubble distance $D_H (z)$ is given by 
\begin{equation}
D_H (z)  =   \frac{1}{ H(z)} \,, 
\end{equation}
where $H(z)$ is the Hubble parameter at the redshift $z$. We note that we adopt the natural unit where $c=\hbar = 1$ in this paper.  The line-of-sight comoving distance $D_C (z)$ is 
\begin{equation}
D_C (z)  =  \int_0^z \frac{1}{H (z') } dz' \,.
\end{equation}
The  (comoving) angular diameter distance $D_M (z)$ is calculated as 
\begin{equation}
D_M (z) = 
\begin{cases}
 \displaystyle\frac{\sin \left[ \sqrt{-\Omega_K} H_0 D_C (z) \right]}{\sqrt{-\Omega_K} H_0 } & {\rm for}~~ \Omega_K<0 ~~{\rm (closed)}  \\ \\
 D_C (z)  &  {\rm for}~~ \Omega_K = 0 ~~{\rm (flat)}   \\ \\
 \displaystyle\frac{\sinh \left[ \sqrt{\Omega_K} H_0 D_C (z) \right]}{\sqrt{\Omega_K} H_0 } & {\rm for}~~ \Omega_K>0 ~~{\rm (open)}  \\ \\ 
\end{cases}
\,,
\end{equation}
where $\Omega_K$ is the density parameter for the curvature of the Universe. The sound horizon at the drag epoch $z_d$ is 
\begin{equation}
r_d = \int_{z_d}^{\infty} \frac{c_s(z)}{H(z)} dz \,,
\end{equation}
where $c_s$ is the sound speed which is given by $c_s = \sqrt{1 /(3 ( 1 + 3 \rho_b  (z) /4 \rho_\gamma (z) ))}$ with $\rho_b (z)$ and $\rho_\gamma (z) $ being the energy densities of baryon and photon at the redshift $z$.\footnote{
Actually, we need the CMB temperature of $T_0 = 2.7255~{\rm K}$ to calculate the energy density of radiation. In this sense, our analysis is not completely free from CMB even though we do not use the data from CMB temperature anisotropies and polarizations. 
} 

Observations of BAO probe the quantities $D_M(z)/r_d$ and $D_H (z) / r_d$ which measure the distances along the transverse and  line-of-sight directions, respectively. In some cases, a spherically-averaged distance is also used, which can be characterized as
\begin{equation}
D_V (z) / r_d  \equiv \left( z D_M^2 (z) D_H(z) \right)^{1/3} /r_d \,.
\end{equation}

In our analysis, we adopt the BAO measurements from SDSS main galaxy sample (MGS), BOSS DR12 galaxies, eBOSS luminous red galaxies (LRGs), eBOSS quasars,  and Lyman $\alpha$ forest  and its cross correlation with quasars from SDSS which are compiled in \cite{eBOSS:2020yzd}.  We summarize these BAO data and their observables in each data set in Table~\ref{tab:BAO}. Although the full likelihood analysis  to derive constraints from these BAO measurements is desirable, we instead use the values tabulated in Table~\ref{tab:BAO} to calculate $\chi^2  \, (\propto 2 \ln {\cal L})$ by neglecting the off-diagonal components of the covariance matrix. We checked that the derived constraints on $H_0$ and $\Omega_m$ are consistent with the ones given in \cite{eBOSS:2020yzd}  and we believe our treatment sufficiently captures constraints from BAO. 

When we adopt BAO data, we also make a separate analysis using the data only from low-$z$ ($z<1$) or high-$z$ ($z>1$) measurement  as has been done in some works \cite{Schoneberg:2019wmt,eBOSS:2020yzd} in addition to the analysis with all redshift data combined. Although we could also make a distinction by using the criterion that the data comes from either galactic  or Lyman $\alpha$, here we follow \cite{eBOSS:2020yzd} to separate the data using the redshift of $z=1$. As we discuss in the next section, low and high redshift data show different tendencies with regard to constraints on $H_0$ and other parameters. 

\begin{table}
\begin{center}
\begin{tabular}{c|l|c|l}
\hline \hline
& Data  & $z_{\rm eff}$  &  observables\\ \hline 
low-$z $
& MGS   &  $0.15$  &  $D_V(z)  / r_d = 4.47 \pm 0.17$ \\ \cline{2-4}
$(z<1)$
& BOSS galaxy I   &  $0.38$  &  $D_M(z)  / r_d = 10.23 \pm 0.17, \, D_H(z) / r_d = 25.00 \pm 0.76$ \\ \cline{2-4}
BAO
& BOSS galaxy II  &  $0.51$  &  $D_M(z)  / r_d = 13.36 \pm 0.21, \, D_H(z) / r_d = 22.33 \pm 0.58$ \\ \cline{2-4}
& eBOSS LRG  &  $0.7$  &  $D_M(z)  / r_d = 17.86 \pm 0.33, \, D_H(z) / r_d = 19.33 \pm 0.58$ \\ \cline{2-4}
& eBOSS ELG  &  $0.85$  &  $D_V(z)  / r_d = 18.33^{+0.57}_{-0.62}$ \\ \hline
high-$z$ 
& eBOSS Quasar  &  $1.48$  &  $D_M(z)  / r_d = 30.69 \pm 0.80, \, D_H(z) / r_d = 13.26 \pm 0.55$ \\ \cline{2-4}
$(z>1)$
& Ly$\alpha$-Ly$\alpha$  &  $2.33$  &  $D_M(z)  / r_d = 37.6 \pm 1.9, \, D_H(z) / r_d = 8.93 \pm 0.28$ \\ \cline{2-4}
BAO
& Ly$\alpha$-Quasar  &  $2.33$  &  $D_M(z)  / r_d = 37.3 \pm 1.7, \, D_H(z) / r_d = 9.08 \pm 0.34$ \\ \cline{2-4}
\hline \hline
\end{tabular}
\caption{BAO data adopted in this paper \cite{eBOSS:2020yzd} \label{tab:BAO}}
\end{center}
\end{table}

Regarding supernovae (SN) data, we adopt the Pantheon sample \cite{Scolnic:2017caz} which includes SNeIa  in the redshift range of $ 0.01 < z <2.3$. We use the likelihood code for SNe provided as a part of {\tt cosmomc} package \cite{Lewis:2002ah}.  As we explain in the following subsection, we consider some dark energy models which allow the time variation of its equation of state as an extension of the $\Lambda$CDM model. In such models, the equation of state parameters can be severely constrained by the SN data, which is quite effective in removing the degeneracy among the  parameters.

We also include the BBN measurements of the helium abundance $Y_p$ \cite{Aver:2015iza} and the deuterium abundances $y_{\rm DP}$ \cite{Cooke:2017cwo} which are respectively given by 
\begin{eqnarray}
Y_p & \left( \equiv \displaystyle\frac{4 n_{\rm He}}{n_b}  \right)=  & 0.2449 \pm 0.0040  ~~(68 ~\% ~{\rm C.L.}) \,, \\ [8pt]
y_{\rm DP}  & \left( \equiv \displaystyle\frac{10^5 \, n_D}{n_b}  \right)= &  2.527 \pm 0.030  ~~(68 ~\% ~{\rm C.L.}) \,,
\end{eqnarray} 
where $n_{\rm He}, n_D$ and $n_b$ are number densities of helium 4, deuterium and baryon.   Since there are  uncertainties coming from several nuclear reaction rates, including the neutron life time,  we also include theoretical uncertainties  for helium abundance as $\sigma_{\rm th} ({\rm He}) = 3.0 \times 10^{-4}$ and  for deuterium as $\sigma_{\rm th} (D) = 0.050$ in the analysis  \cite{Coc:2015bhi,Consiglio:2017pot}.  We use a public code {\tt PArthENoPE} \cite{Consiglio:2017pot} to calculate the theoretical values of $Y_p$ and $y_{\rm DP}$ for a given cosmological model. The deuterium abundance can severely constrain baryon density which cannot be well determined by BAO and SN. When the effective number of neutrinos $N_{\rm eff}$ is allowed to vary, it can be strongly constrained by the helium abundance data.

By including the data from BAO, SN and BBN, we derive constraints on $H_0$ as well as other model parameters by performing MCMC analysis.  Convergence of a chain is diagnosed based on the integrated auto correlation time following \cite{10.2140/camcos.2010.5.65,2013PASP..125..306F}. More concretely, we stop MCMC when the chain length is at least one hundred times larger than any of the autocorrelation times of the primary parameters.  We discard the first half of the chains as burn-in.

\subsection{Models to be analyzed}

In several works, it has been argued that the $H_0$ tension between direct and indirect measurements exists even without CMB data in the $\Lambda$CDM and some simple extended models \cite{Addison:2017fdm,Abbott:2017smn,Cuceu:2019for,Schoneberg:2019wmt,eBOSS:2020yzd,Philcox:2020vvt}.  In this paper, we investigate this issue further by considering a range of  extended models using the data from BAO, SN and BBN, i.e., without CMB. The models considered in this paper are summarized in Table~\ref{tab:models} along with the parameters in each model and the prior ranges of them are tabulated in Table~\ref{tab:prior}.  Below we describe each model in some detail.

\begin{table}
\begin{center}
\begin{tabular}{l|l}
\hline \hline
model  &  model parameters  \\ \hline 
$\Lambda$CDM   &  $\omega_b, \, \omega_{\rm dm}, \omega_{\Lambda}$   \\
$w$CDM   &  $\omega_b, \, \omega_{\rm dm}, \omega_{\rm DE},  w$   \\
$N_{\rm eff} \Lambda$CDM   & $\omega_b, \, \omega_{\rm dm}, \omega_{\Lambda}, N_{\rm eff}$  \\
$\Omega_K \Lambda$CDM  & $\omega_b, \, \omega_{\rm dm}, \omega_{\Lambda}, \omega_K$   \\
$w_0 w_a$(CPL)CDM  & $\omega_b, \, \omega_{\rm dm}, \omega_{\rm DE}, w_0, w_a$    \\
$N_{\rm eff}  \Omega_K \Lambda$CDM & $\omega_b, \, \omega_{\rm dm}, \omega_{\Lambda}, N_{\rm eff}, \omega_K$    \\
$N_{\rm eff}w_0 w_a $(CPL)CDM   & $\omega_b, \, \omega_{\rm dm}, \omega_{\rm DE}, N_{\rm eff}, w_0, w_a$   \\
$w_0 w_1 w_2$CDM (binned DE EoS)  & $\omega_b, \, \omega_{\rm dm}, \omega_{\rm DE}, w_0, w_1, w_2$    \\
\hline \hline
\end{tabular}
\caption{Models analyzed in this paper and the model parameters. \label{tab:models}}
\end{center}
\end{table}

\begin{table}
\begin{center}
\begin{tabular}{l|l}
\hline \hline
parameter  &  prior range   \\ \hline 
$\omega_b$   & $[0.001, 0.1]$    \\
$\omega_{\rm DM}$    & $[0.01,  0.3]$  \\
$\omega_\Lambda (\omega_{\rm DE}) $    & $[0.1, 0.5]$   \\
$\omega_K$    & $[-0.3,  0.3]$   \\
$w (w{\rm CDM}) $  & $[-4,  1]$    \\
$w_0 {\rm (CPL)} $   & $[-3,  -0.5]$    \\
$w_a {\rm (CPL)}$  & $[-3,  2]$   \\
$w_0 {\rm (binned)}$    & $[-3,  0]$    \\
$w_1 {\rm (binned)}$    & $[-3,  0]$    \\
$w_2 {\rm (binned)}$    & $[-3,  0]$   \\
$N_{\rm eff}$    & $[1, 5]$    \\
\hline \hline
\end{tabular}
\caption{Prior range of model parameters adopted in the analysis. \label{tab:prior}}
\end{center}
\end{table}

\bigskip
\noindent 
$\rm{a}.$ {\bf $\bm{\Lambda}$CDM model}  \\
Since we use the data from BAO, SN and BBN, which only probe the background evolution, we just vary $\omega_b \left( = \Omega_b h^2 \right), \, \omega_{\rm DM} \left( = \Omega_{\rm DM} h^2 \right)$ and  $\omega_\Lambda \left( = \Omega_\Lambda h^2 \right)$ where $\Omega_b, \Omega_{\rm DM}$ and $\Omega_\Lambda$ are the density parameters for baryon, dark matter and the cosmological constant, respectively. $h$ is the reduced Hubble constant (i.e., $H_0$ in units of $100 ~{\rm km/s/Mpc}$). Since a flat Universe is assumed in this model,  $h$ can be derived by $h = \sqrt{\omega_m + \omega_\Lambda}$ with $\omega_m = \omega_b + \omega_{\rm DM}$.

\bigskip
\noindent 
$\rm{b}.$ {\bf $\bm{w}$CDM model} \\
In this model, we also vary the equation of state (EoS) of dark energy which is assumed to be constant in time and denoted as $w$. Therefore the free parameters in this model are $(\omega_b, \omega_{\rm DM}, \omega_{\rm DE}, w)$ where $\omega_{\rm DE}$ is the density parameter for dark energy (equivalent to $\omega_\Lambda$).

\bigskip
\noindent 
$\rm{c}.$ {\bf $\bm{N_{\rm eff} \Lambda}$CDM model} \\
In the $\Lambda$CDM and its other extended models investigated in this paper, the effective number of neutrinos is fixed as $N_{\rm eff} = 3.046$ unless otherwise stated.  Since extra (or less) effective neutrino numbers directly affect BBN and the sound horizon at the drag epoch (which can also modify the fit to BAO data) we also consider an extension of $\Lambda$CDM by adding $N_{\rm eff}$ as a free parameter.  A cosmological constant  is assumed for dark energy and the  parameters in this model are $(\omega_b, \omega_{\rm DM}, \omega_\Lambda, N_{\rm eff})$.

\bigskip
\noindent 
$\rm{d}.$ {\bf $\bm{\Omega_K \Lambda}$CDM model} \\
We also consider an extension of $\Lambda$CDM model by allowing the curvature of the Universe to be varied by adding the parameter  $\omega_K (=\Omega_K h^2)$ in the analysis. This model just assumes a nonflat Universe and others are the same as the ones in $\Lambda$CDM case.  The reduced Hubble constant $h$ is given by $h = \sqrt{\omega_m + \omega_\Lambda + \omega_K}$ in this model. The  parameters in this framework are $(\omega_b, \omega_{\rm DM}, \omega_\Lambda, \omega_K)$.

\bigskip
\noindent 
$\rm{e}.$ {\bf $\bm{w_0 w_a}$(CPL)CDM model} \\
This model assumes a time varying EoS for dark energy in the following form:
\begin{equation}
\label{eq:CPL}
w (z) = w_0 + w_a (1-a) = w_0 + \frac{z}{1+z} w_a \,,
\end{equation}
which is the so-called CPL parametrization \cite{Chevallier:2000qy,Linder:2002et}. The  parameters in this model are $(\omega_b, \omega_{\rm DM}, \omega_{\rm DE}, w_0, w_a)$

\bigskip
\noindent 
$\rm{f}.$ {\bf $\bm{N_{\rm eff} \Omega_K \Lambda}$CDM model} \\
In this model, the effective number of neutrinos and the curvature of the Universe are simultaneously varied in addition to the standard cosmological parameters in the $\Lambda$CDM model. The parameters in this model are $(\omega_b, \omega_{\rm DM}, \omega_\Lambda, N_{\rm eff},  \omega_K)$.

\bigskip
\noindent 
$\rm{g}.$ {\bf $\bm{N_{\rm eff} w_0 w_a}$(CPL)CDM model} \\
In this model, the CPL parametrization [Eq.~\eqref{eq:CPL}] for dark energy EoS is adopted and $N_{\rm eff}$ is also varied. Other parameters are the same as the ones in the $\Lambda$CDM model and hence the parameters in this model are $(\omega_b, \omega_{\rm DM}, \omega_{\rm DE}, N_{\rm eff},  w_0, w_a)$

\bigskip
\noindent 
$\rm{h}.$ {\bf $\bm{w_0 w_1 w_2}$CDM model (binned parametrization for dark energy EoS)} \\
We also consider another model for dark energy EoS where different EoS for each linearly binned scale factor range are assumed. We respectively assign the EoS as $w_0, w_1$ and $w_2$ for the scale factor range of $a = \left[1, \frac23 \right], \left[ \frac23, \frac13 \right]$ and $ \left[\frac13 , 0 \right]$, which corresponds to the redshift bins of $z = [0, 0.5], [0.5, 2]$ and $ [2, \infty]$  (i.e., $w_0$ is the EoS for the redshift range of $0< z < 0.5$,  $w_1$ is for the range of $ 0.5 < z < 2$ and $w_2$ is for the redshift of $z>2$).  The  parameters in this model are $(\omega_b, \omega_{\rm DM}, \omega_{\rm DE}, w_0, w_1, w_2)$.

\bigskip\bigskip

We investigate whether the $H_0$ tension  still persists (or not) even without CMB data in the $\Lambda$CDM  and seven extended models  described above.  In the next section, we present the results of our analysis  for each model in order.

\section{Constraints on $H_0$ without CMB  \label{sec:results}}

\begin{figure}
\centering
\scalebox{0.65}{\includegraphics{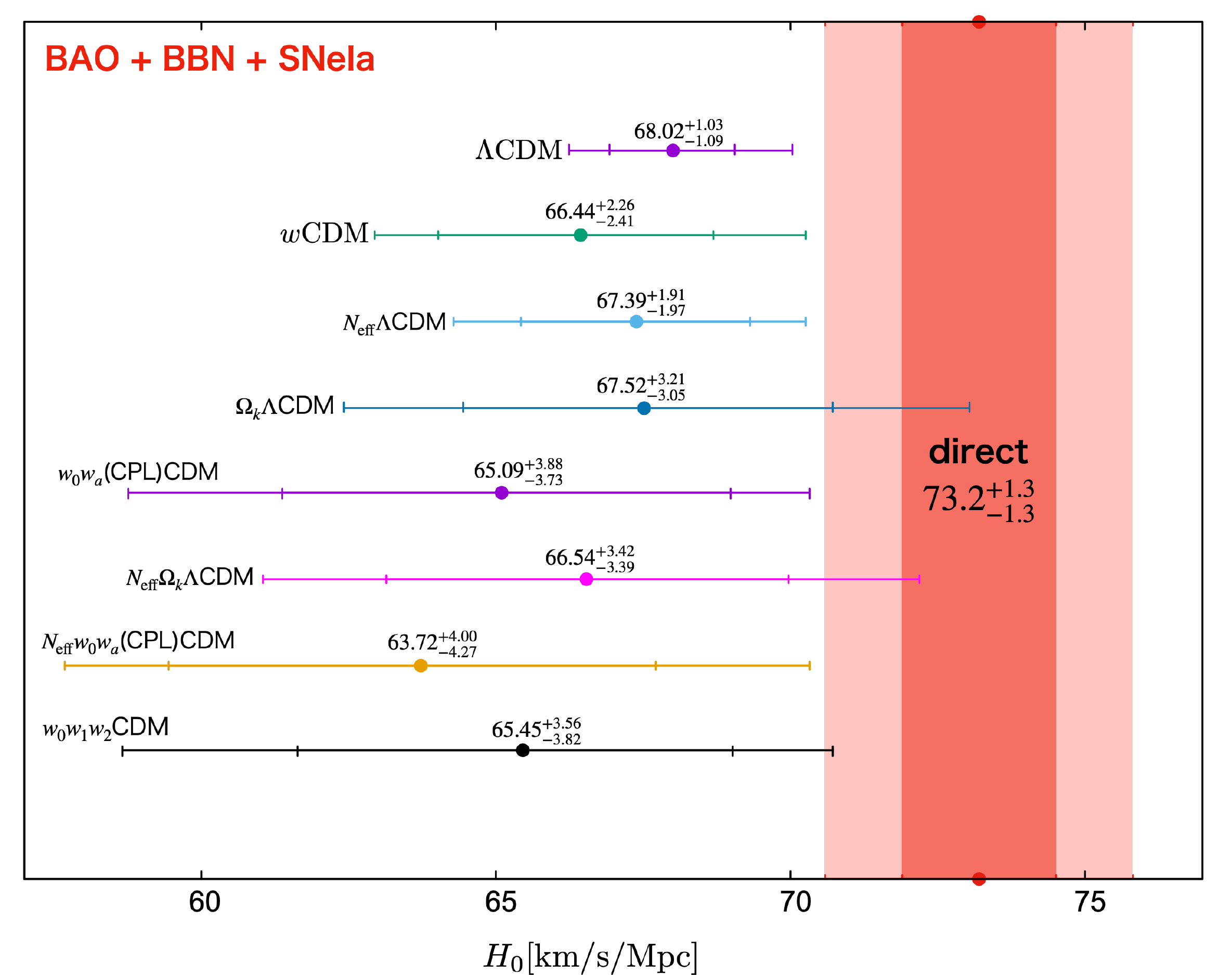}}
\caption{Summary of constraints on $H_0$ from BAO+BBN+SN in $\Lambda$CDM and the extended models. The mean value is represented by a filled circle and 1$\sigma$ and 2$\sigma$ ranges for each model are indicated  by short vertical bars. The value of $H_0$ obtained by the local measurement from SH0ES, $ H_0 = 73.2 \pm 1.3$ \cite{Riess:2020fzl}, is also shown with vertical bands (the inner and outer ones corresponds to 1$\sigma$ and 2$\sigma$ C.L.). \label{fig:summary} }
\end{figure}

In this section, we present our results on the constraints for $H_0$ from BAO+SN+BBN in the framework of eight models ($\Lambda$CDM + seven extended models)  described in the previous section.  To discuss whether the $H_0$ tension persists in those models, we show a 2D constraint in the $\Omega_m$-$H_0$ plane and their 1D posterior distribution. In addition, we also show 2D constraints and 1D posterior distributions for the parameters beyond the $\Lambda$CDM model in the following.  The summary of the constraint on $H_0$ in each model is given in Fig.~\ref{fig:summary}. Below we discuss constraints for each model in order.

\begin{figure}
\centering
\scalebox{0.65}{\includegraphics{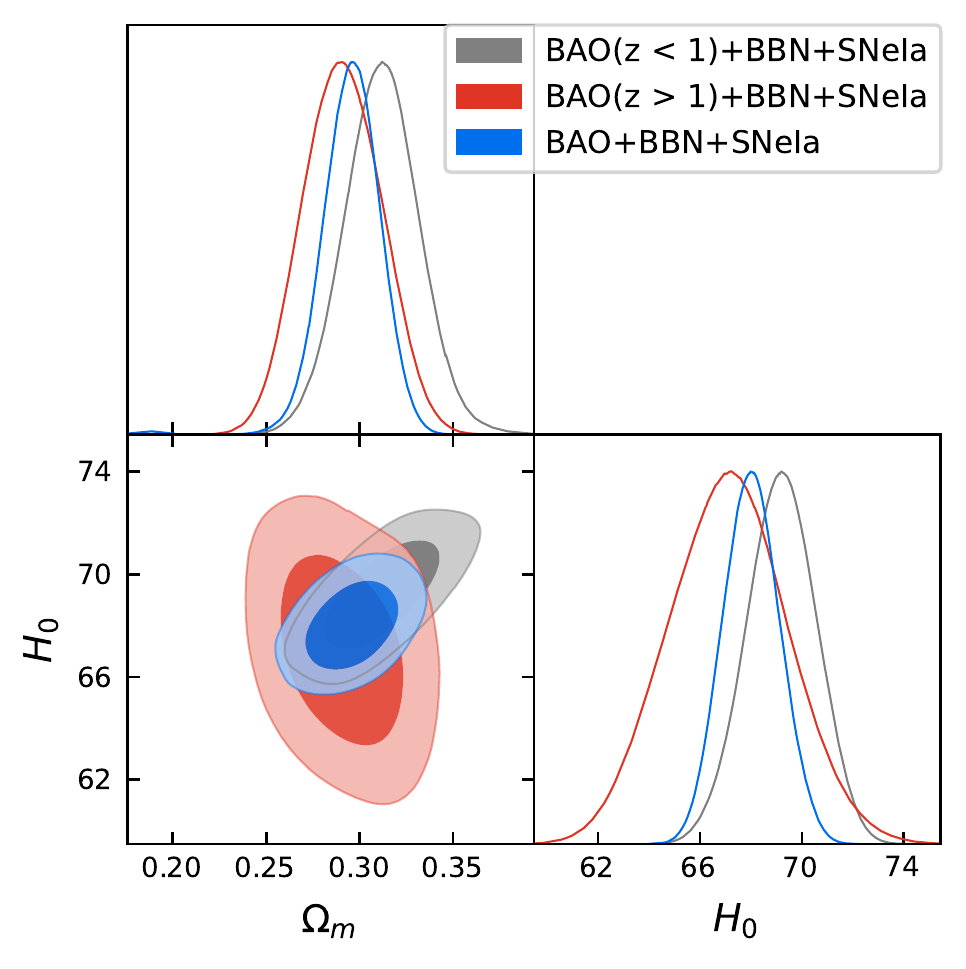}}
\caption{Constraints for $\Lambda$CDM model \label{fig:LCDM} }
\end{figure}

\bigskip\bigskip
\noindent 
{\bf $\bm \Lambda$CDM model}: 
First we start from the $\Lambda$CDM model, in which constraints are shown in Fig.~\ref{fig:LCDM}.  As already studied in \cite{Addison:2017fdm,Abbott:2017smn,Cuceu:2019for,Schoneberg:2019wmt,eBOSS:2020yzd,Philcox:2020vvt}, the $H_0$ tension  exists even without CMB data, which we also confirm from our analysis. It should also be noted that although low-$z$ ($z<1$) BAO data is in favor of a relatively large value of $H_0$,  the data from high-$z$ ($z>1$)  BAO in contrast prefers a low value of $H_0$.  When we combine low-$z$ and high-$z$ BAO data, their combination gives 
\begin{equation}
\label{eq:LCDM_H0_Om_error}
H_0 = 68.02 ^{+1.03}_{-1.09} \,,
\qquad 
\Omega_m = 0.295 ^{+0.014}_{-0.014}
\qquad
[\Lambda{\rm CDM}] \,, 
\end{equation}
(the uncertainties are quoted for 1$\sigma$) which is significantly lower than the one obtained from the local measurements.  We quote the value of $H_0$ from SH0ES \cite{Riess:2020fzl} 
\begin{equation}
\label{eq:local_H0}
 H_0 = 73.2 \pm 1.3 
\qquad [{\rm SH0ES}]
\end{equation}
as the reference value for the local measurements. Given this value, the significance of the tension between the local (direct) measurements and indirect one from BAO+BBN+SNe (i.e., without CMB) in $\Lambda$CDM model is $3.1\sigma$. 

\begin{figure}
\centering
\scalebox{0.5}{\includegraphics{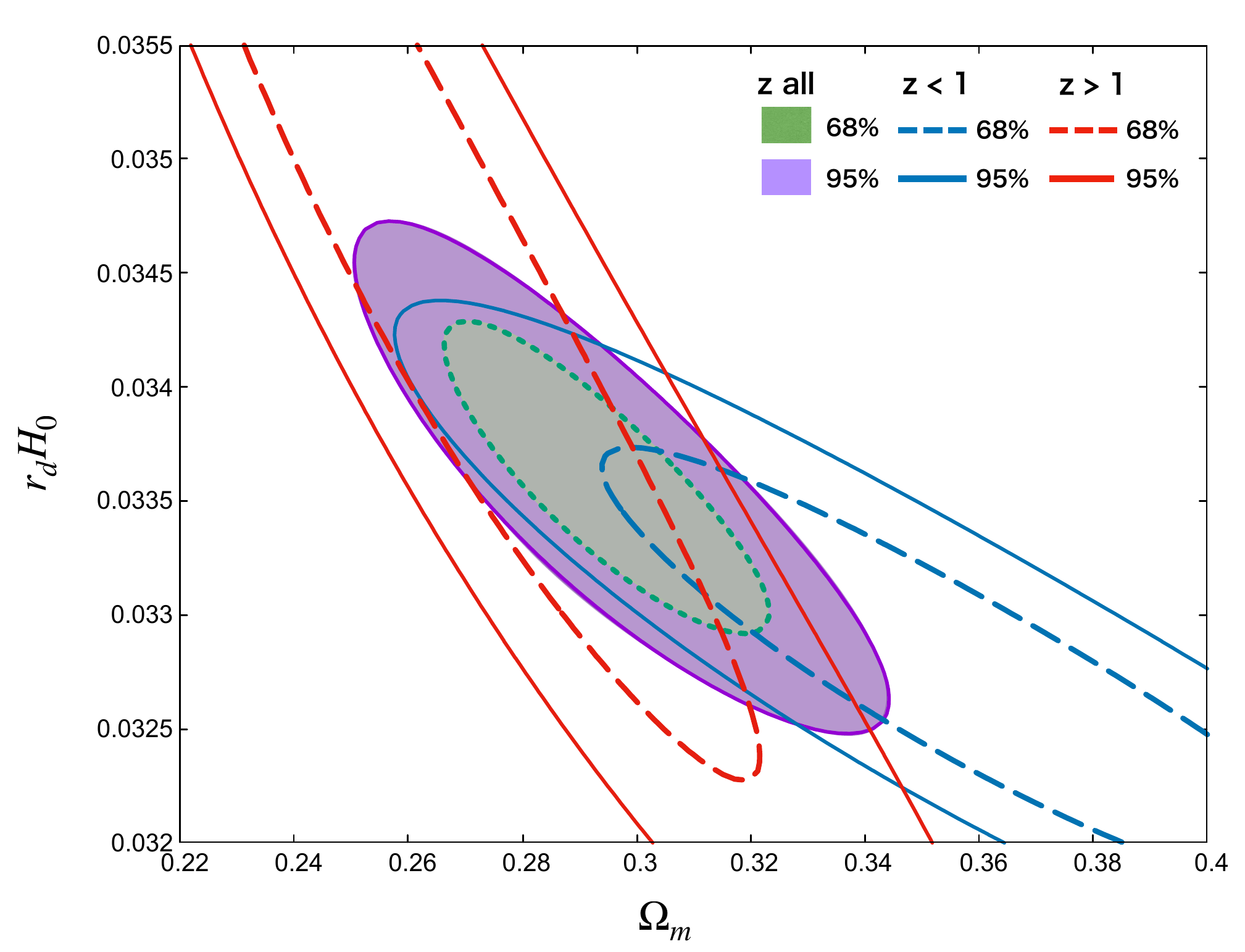}}
\caption{Constraints in the $\Omega_m$--$r_d H_0$ plane for $\Lambda$CDM model from BAO data only. Constraints from low redshift $(z<1)$, high redshift $(z>1)$ and all BAO data are separately shown.  \label{fig:LCDM_BAO} }
\end{figure}

As seen from the constraint in the $H_0$--$\Omega_m$ plane of Fig.~\ref{fig:LCDM},  the direction of the degeneracy between $H_0$ and $\Omega_m$ are different for low-$z$ and high-$z$  BAO data, and hence once they are combined, only the overlapping region is allowed and the constraint on  $H_0$  becomes tight and a low value is preferred.  To see why the directions of the degeneracy are different, we also show constraints in the $\Lambda$CDM model only from BAO data in Fig.~\ref{fig:LCDM_BAO}.  As described in the previous section, the observables of BAO are $D_H(z)/r_d, \, D_M (z) / r_d$ and $D_V (z) /r_d $ which can be  written as 
\begin{eqnarray}
\label{eq:BAO_DH}
\frac{D_H(z)}{r_d}  & = &   \frac{f(z)}{r_d H_0 }  \,, \\ [8pt]
\label{eq:BAO_DM}
\frac{D_M(z)}{r_d}  & = &  \frac{\displaystyle\int_0^z d\bar{z} \,  f(\bar{z})}{r_d H_0 }  \,,  \\ [8pt]
\label{eq:BAO_DV}
\frac{D_V(z)}{r_d}  & = &  \frac{\left[ z \left( \displaystyle\int_0^z d \bar{z}\, f(\bar{z}) \right)^2 \, f(z) \right]^{1/3} }{r_d H_0} \,,
\end{eqnarray}
where $f(z)\equiv H_0/H(z)$ and in a flat $\Lambda$CDM model it is given by
\begin{equation}
\label{eq:f_z_LCDM}
\left. f(z) \right|_{\Lambda{\rm CDM}} = \frac{1}{\sqrt{\Omega_m (1+z)^3 + 1 - \Omega_m }} \,.
\end{equation}
As can be seen from Eqs.~\eqref{eq:BAO_DH}--\eqref{eq:f_z_LCDM},  the BAO  observables for a given redshift only depend on the combination of $r_d H_0$ and $\Omega_m$ in a  flat $\Lambda$CDM model \cite{Addison:2013haa}. Although the primary parameters adopted in the analysis are $\omega_b$, $\omega_{\rm DE}$ and $\omega_\Lambda$, to obtain a simple understanding of how the direction of the degeneracy appears for low and high redshift BAO data, we show the constraint in the $\Omega_m$--$r_d H_0$ plane in Fig.~\ref{fig:LCDM_BAO}. Here it should be cautioned that in the analysis of Fig.~\ref{fig:LCDM_BAO},  $\Omega_m$ and $r_d H_0$ are treated as independent variables although $r_d$ actually also depends on $\Omega_m$ and its dependence is automatically incorporated  in the main analysis  whose  result is shown in Fig.~\ref{fig:LCDM}.

Regarding $D_H(z) / r_d$,   it  depends on $r_d H_0$ and $\Omega_m$ at high redshifts ($z>1)$ approximately as
\begin{equation}
\label{eq:rd_H0_Om}
\frac{D_H(z)}{r_d} \propto \frac{1}{r_d H_0 \sqrt{\Omega_m}} \,,
\end{equation}
since a term with $(1+z)^3$ dominates  in the denominator of Eq.~\eqref{eq:BAO_DH}. This  indicates that $r_dH_0$ and $\Omega_m$ are degenerate along this direction. On the other hand, at low redshift ($z<1$), the function $f(z)|_{\Lambda{\rm CDM}}$ can be approximated by some leading terms of the expansion at $z=0$ as 
\begin{equation}
\left. f(z) \right|_{\Lambda{\rm CDM}} = 1 - \frac32 \Omega_m z + {\cal O}(z^2) \,,
\end{equation}
from which one can see that the $\Omega_m$ dependence becomes weak for lower redshifts and hence the constraints on $\Omega_m$ get less severe when the low-$z$ BAO data is considered. The tendency is also the same for $D_M(z) /r_d$ in which the function $f(z)|_{\Lambda{\rm CDM}}$ is integrated over $z$. Therefore the direction of the degeneracy in the $\Omega_m$--$r_d H_0$ plane gets inclined parallel to the $\Omega_m$ axis when low-$z$ data is considered.  From this reasoning, the allowed regions  from low-$z$ ($z<1$) and high-$z$ ($z>1$) BAO data extend to different directions as seen from Fig.~\ref{fig:LCDM_BAO}. However, when low and high redshift data are combined, only the overlapping region is allowed.  Although we have discussed the $\Omega_m$--$r_d H_0$ plane here, the same argument should also apply to the constraint in the $H_0$--$\Omega_m$ plane, which  explains why the directions of the degeneracies differ between the constraints from low-$z$ and high-$z$ BAO data.

\begin{figure}
\centering
\scalebox{0.65}{\includegraphics{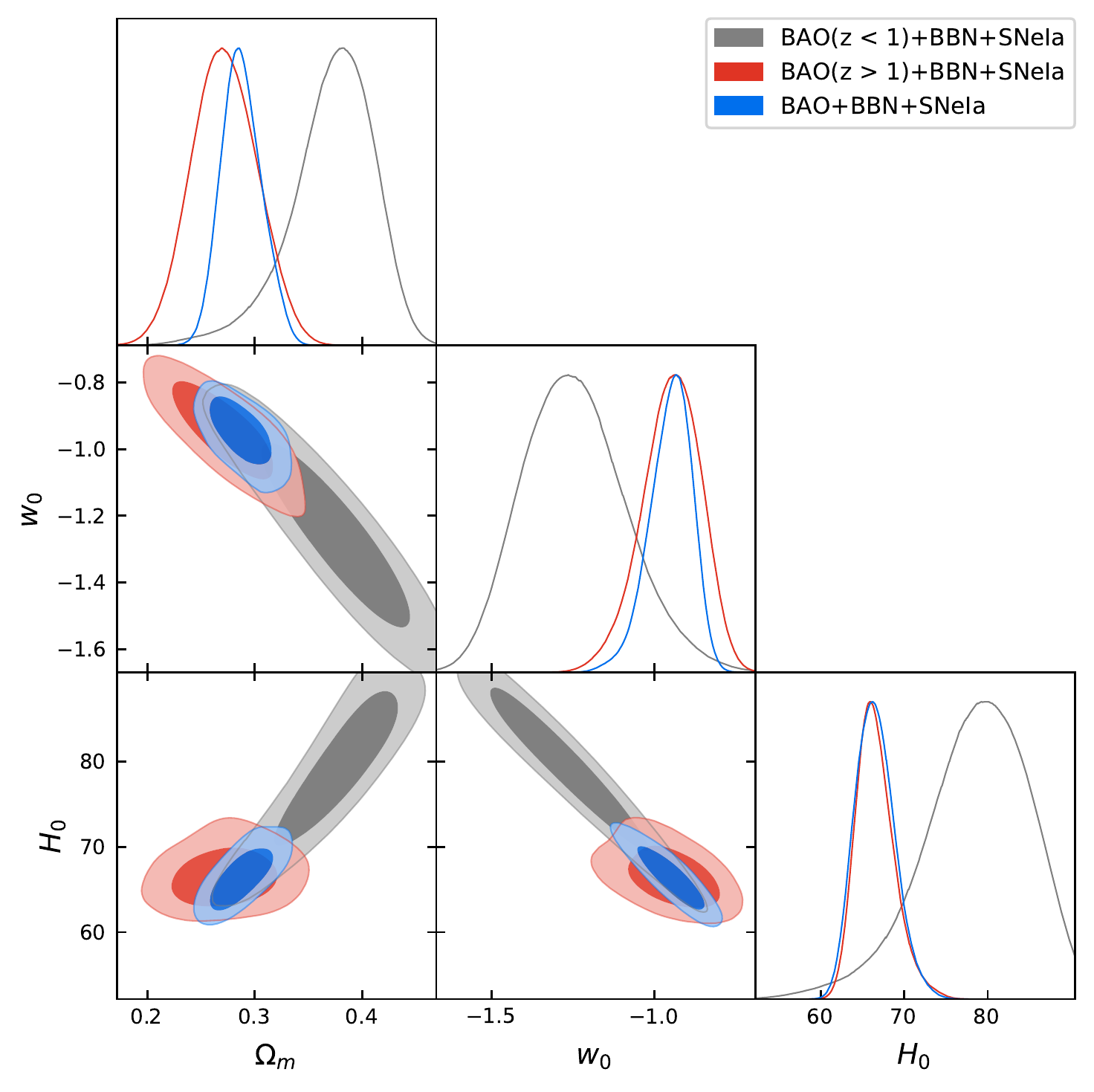}}
\caption{Constraints for $w$CDM model   \label{fig:wCDM} }
\end{figure}

\begin{figure}
\centering
\scalebox{0.5}{\includegraphics{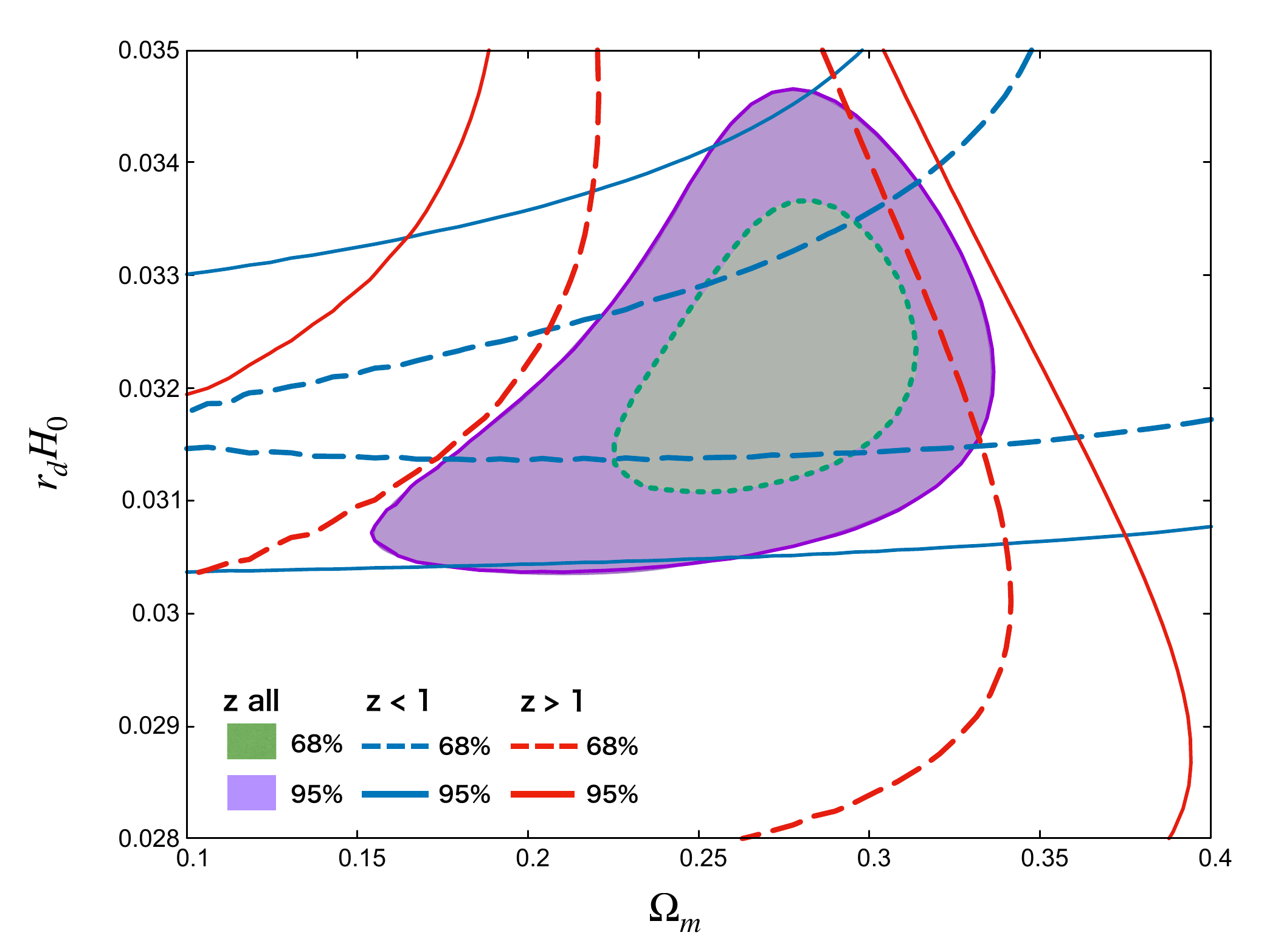}}
\caption{Constraints in the $\Omega_m$--$r_d H_0$ plane for $w$CDM model from BAO data only. Constraints from low redshift $(z<1)$, high redshift $(z>1)$ and all BAO data are separately shown.  \label{fig:wCDM_BAO} }
\end{figure}

\bigskip\bigskip
\noindent 
{\bf $\bm w$CDM model}:   In the $w$CDM model, the function $f(z)$ is given by 
\begin{equation}
\left. f  (z) \right|_{w{\rm CDM}} = \frac{1}{\sqrt{\Omega_m (1+z)^3 + (1 - \Omega_m)(1+z)^{3(1+w)} }} \,.
\end{equation}
At low-$z$ redshift, this can be expanded as 
\begin{equation}
\left. f  (z) \right|_{w{\rm CDM}} = 1   + \frac32 \left( -1 - w + w \Omega_m \right) z + {\cal O}(z^2) \,.
\end{equation}
Therefore, as in the case for the $\Lambda$CDM model, the low-$z$ BAO data cannot severely constrain $\Omega_m$, which explains the reason why the allowed region for low-$z$ BAO data extends to a higher value of $\Omega_m$ in the $\Omega_m$--$H_0$ and $\Omega_m$--$w$ planes in Fig.~\ref{fig:wCDM}. This can also be understood from the constraints in the $\Omega_m$--$r_d H_0$ plane, which is shown in Fig.~\ref{fig:wCDM_BAO}.  On the other hand, high-$z$ BAO data can constrain $\Omega_m$ well although there is degeneracy between $r_d H_0$ and $\Omega_m$ as can be understood from Eq.~\eqref{eq:rd_H0_Om}. There also exists a degeneracy between $\Omega_m$ and $w$, however, as the redshift increases, the nature of dark energy does not affect the fit to BAO data much, and hence,  $\Omega_m$ can be relatively determined from high-$z$ data. This can be seen from  Fig.~\ref{fig:wCDM_BAO}. 

Regarding $w$,  $\left. f  (z) \right|_{w{\rm CDM}}$ at low-$z$ does not depend on $w$ at leading order, and hence $w$ cannot be well constrained  by low-$z$ data, which is also found in the  2D constraints shown in the panel of the $\Omega_m$--$w$ and $H_0$--$w$ plane in Fig.~\ref{fig:wCDM}. Since $w$ and $H_0$ are degenerate in $D_H(z)$ and $D_M(z)$, when a broad range of $w$ is allowed, $H_0$ would also take a wide range of values along the direction of the degeneracy. In particular, as more negative values of $w$ is allowed from low-$z$ data, higher $H_0$ can also be tolerated as can be seen from Fig.~\ref{fig:wCDM}.  However, high-$z$ BAO data can severely constrain $\Omega_m$ and $w$ is limited to be close to $w = -1$ from SN data; hence eventually the constraints for $H_0$ and $\Omega_m$ from the combination of all BAO data are driven to 
\begin{equation}
\label{eq:wCDM_H0_Om}
H_0 = 66.44 ^{+2.26}_{-2.41}  \,,
\qquad
\Omega_m = 0.288 ^{+0.020}_{-0.018}
\qquad
[w{\rm CDM}] \,,
\end{equation}
which is consistent with the one obtained by Planck. The tension in $H_0$ with local direct measurements still persists although the uncertainty for $H_0$ is larger than that in $\Lambda$CDM case due to the degeneracy of $w$ with other parameters. Therefore the  significance in this model is reduced to 2.5$\sigma$.

\begin{figure}
\centering
\scalebox{0.65}{\includegraphics{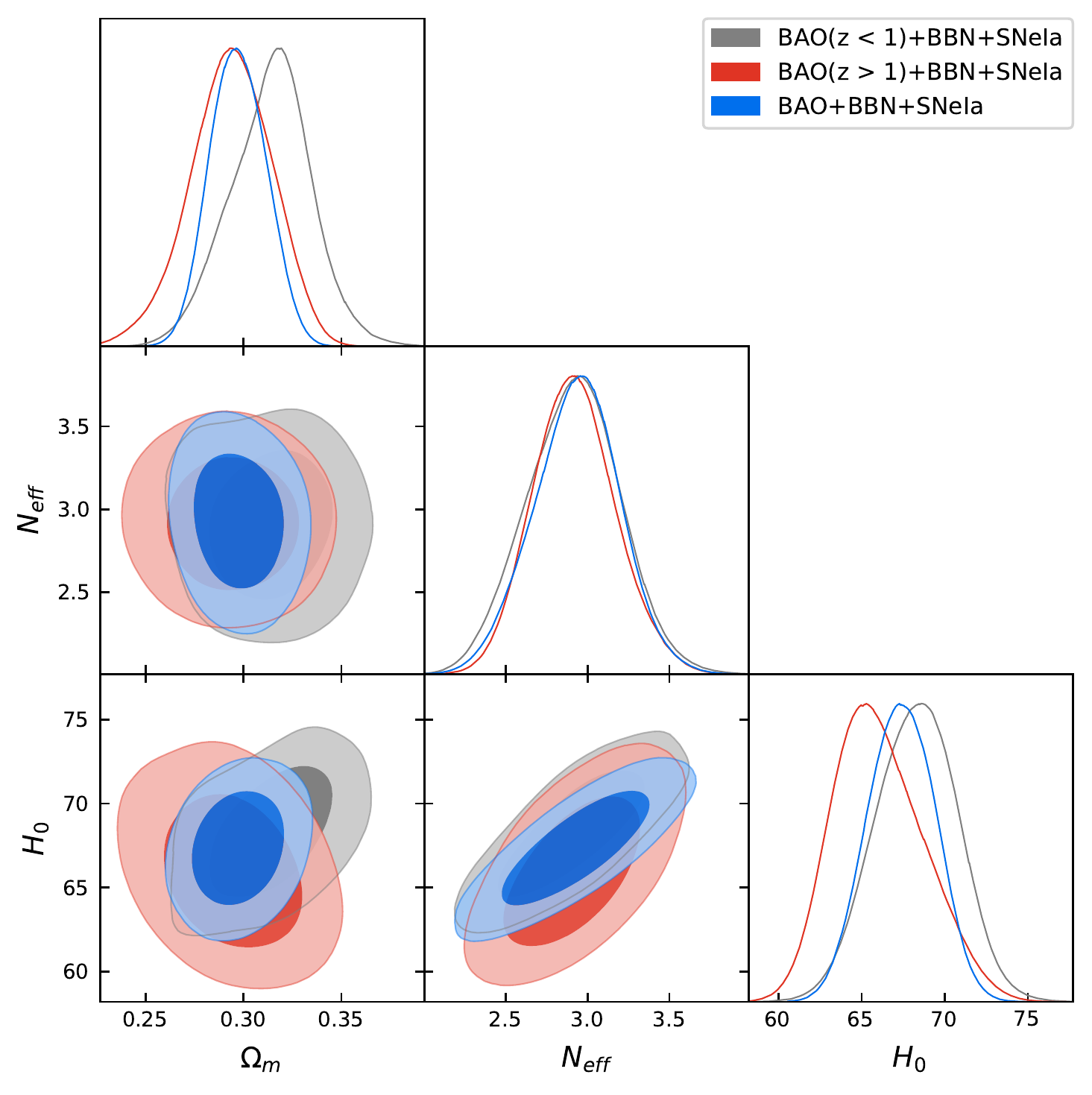}}
\caption{Constraints for $N_{\rm eff} \Lambda$CDM model  \label{fig:NeffLCDM} }
\end{figure}

\bigskip\bigskip
\noindent 
{\bf $\bm{N_{\rm eff} \Lambda}$CDM model}:  In Fig.~\ref{fig:NeffLCDM}, the constraints for $\Omega_m, N_{\rm eff}$ and $H_0$ are shown.  When $N_{\rm eff}$ is allowed to vary, the sound horizon $r_d$ is mainly changed, while the quantities relevant to the evolution after recombination are almost unaffected. Therefore the evolution at low redshift is virtually the same as the one for  the $\Lambda$CDM model. 

However, as seen from Fig.~\ref{fig:NeffLCDM} (due to the degeneracy between $N_{\rm eff}$ and $H_0$ in the sound horizon) the uncertainty of $H_0$ gets larger compared to that for the $\Lambda$CDM case. Actually, we also include BBN data in our analysis and the helium abundance in particular can place a severe constraint on $N_{\rm eff}$, which partially removes the degeneracy between $N_{\rm eff}$ and $H_0$. However, the uncertainty  for $H_0$ gets larger  in this framework too and the constraints for $H_0$ and $\Omega_m$ are given as 
\begin{equation}
\label{eq:NeffCDM_H0_Om}
H_0 = 67.39 ^{+1.91}_{-1.97} \,,
\qquad 
\Omega_m = 0.297 ^{+0.013}_{-0.015} 
\qquad
[N_{\rm eff}\Lambda{\rm CDM}] \,.
\end{equation}
Although the mean value of $H_0$ is slightly lower than that in the $\Lambda$CDM model, its uncertainty gets larger. As a result, the $H_0$ tension, with the local measurements,  is 2.5$\sigma$ in this model.

\begin{figure}
\centering
\scalebox{0.65}{\includegraphics{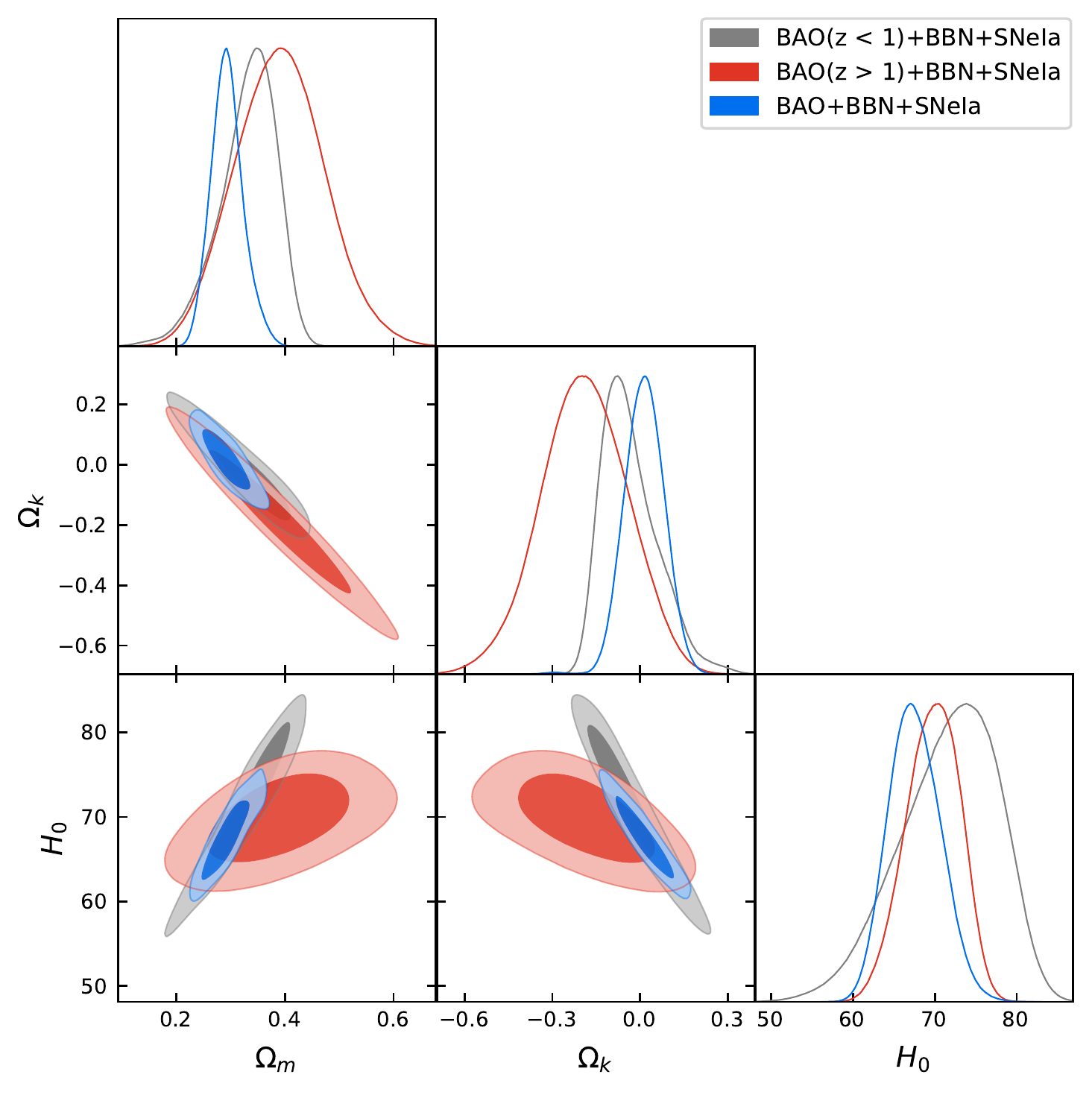}}
\caption{Constraints for $\Omega_K \Lambda$CDM model \label{fig:OkLCDM} }
\end{figure}

\begin{figure}
\centering
\scalebox{0.4}{\includegraphics{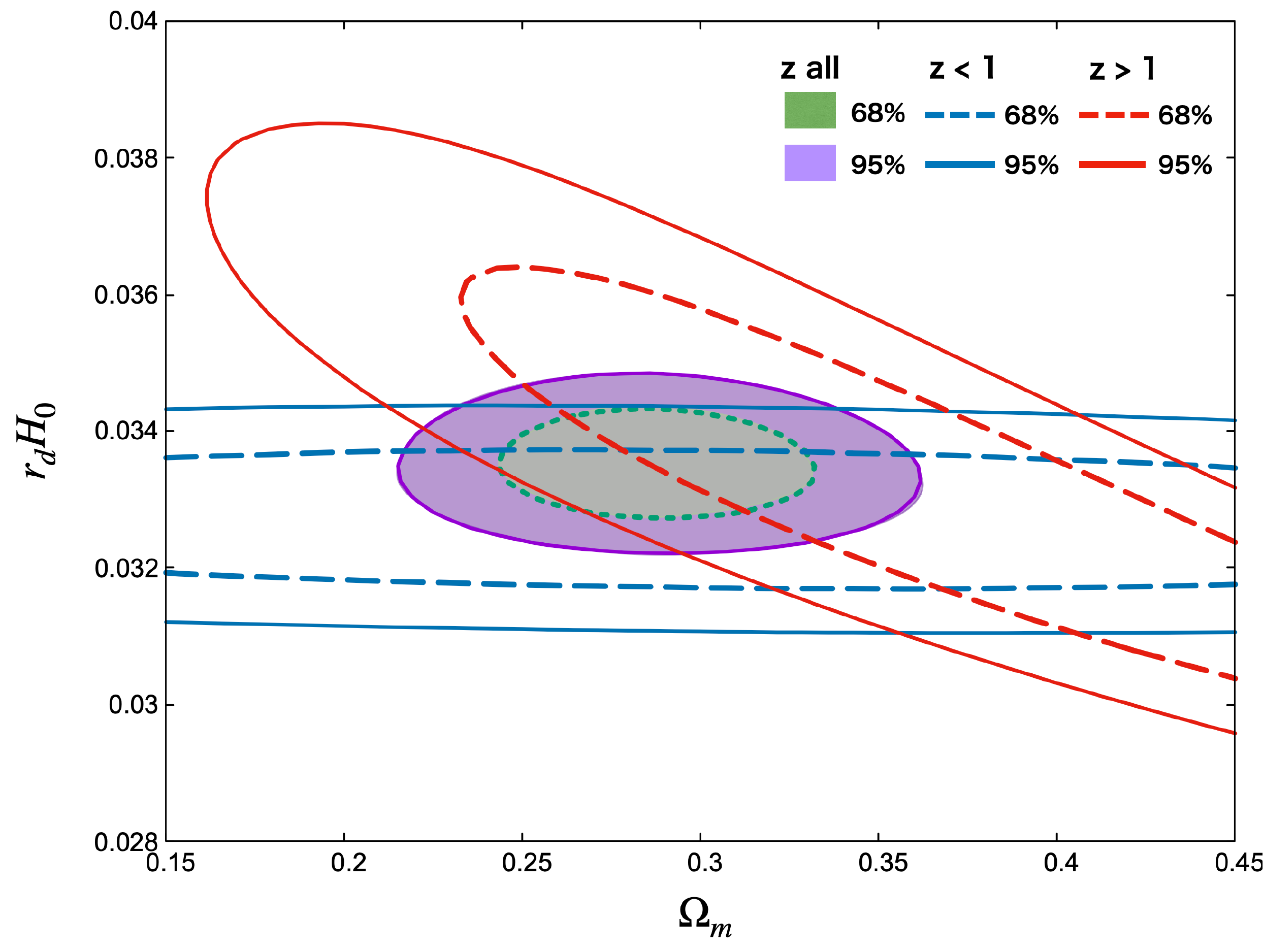}}
\caption{Constraints in the $\Omega_m$--$r_d H_0$ plane for $\Omega_K$CDM model from BAO data only. Constraints from low redshift $(z<1)$, high redshift $(z>1)$ and all BAO data are separately shown.  \label{fig:OkCDM_BAO} }
\end{figure}

\bigskip\bigskip
\noindent 
{\bf $\bm{\Omega_K \Lambda}$CDM model}:  Constraints for $\Omega_m, \Omega_K$ and $H_0$ in this model are presented in Fig.~\ref{fig:OkLCDM}. We also show Fig.~\ref{fig:OkCDM_BAO} where the constraints only from BAO data are depicted in the $\Omega_m$--$r_d H_0$ plane, to understand how the parameters are limited in this framework. The function $f(z) $ in $\Omega_K \Lambda$CDM model is given by
\begin{equation}
\left. f  (z) \right|_{\Omega_K {\rm CDM}} = 
\frac{1}{\sqrt{\Omega_m (1+z)^3 + (1 - \Omega_m-\Omega_K)  + \Omega_K (1+z)^2}} \,.
\end{equation}
Expanding $\left. f(z) \right|_{\Omega_K {\rm CDM}} $ at $z=0$, one obtains
\begin{eqnarray}
\label{eq:low_z_exp_OkLCDM}
\left. f  (z) \right|_{\Omega_K {\rm CDM}} &=& 
1-\frac{3\Omega_m+2\Omega_K}{2}z+{\cal O}\left(z^2\right) \,.
\end{eqnarray}
The dependence on $\Omega_m$ and $\Omega_K$  appears only from the first order in $z$,  and hence low-$z$ BAO data is not sensitive to $\Omega_K$ and $\Omega_m$ in the function $\left. f(z) \right|_{\Omega_K {\rm CDM}} $. However, it should be noted that $r_d H_0$ actually depends on $\Omega_m$ and $H_0$, which can also constrain $\Omega_m$ indirectly through this dependence. 

On the other hand, at high-$z$, when $| \Omega_K | = {\cal O}(0.1)$,  the curvature energy density can give a sizable contribution to $H(z)$ before the cosmological constant dominates the Universe; hence $\Omega_m$ and $\Omega_K$ are degenerate in $H(z)$, which gives more uncertainty in the determination of $\Omega_m$. This is the reason why the error of $\Omega_m$ in the case of high-$z$ BAO data is larger than the one for the low-$z$ BAO case.

In  Fig.~\ref{fig:OkCDM_BAO}, we also show the constraint in the $\Omega_m$--$r_d H_0$ plane only from BAO data. As seen from the figure, $r_d H_0$ can be well determined from low-$z$ BAO  data; however this means that the degeneracy between $H_0$ and other parameters determining $r_d$ arises and eventually the low-$z$ BAO data allows a broad range of $H_0$ (as can also be read off from Fig.~\ref{fig:OkLCDM}). Even though the allowed region from low-$z$ BAO data in Fig.~\ref{fig:OkCDM_BAO}  extends  horizontally and $\Omega_m$ does not seem to be well constrained by low-$z$ BAO data,  we again emphasize that $\Omega_m$ can be indirectly probed through $r_d H_0$, which is the reason why $\Omega_m$ is relatively well constrained by low-$z$ BAO data.

Regarding the allowed range for $H_0$ (as seen from Fig.~\ref{fig:OkLCDM}, even though a higher $H_0$ value is preferred when low-$z$ ($z<1$) or high-$z$ ($z>1$) BAO data are  separately used)  the combination of all BAO data gives a lower value for $H_0$. This is because  the direction of the degeneracy in the $\Omega_K$--$H_0$ and $\Omega_m$--$H_0$ planes are different for low-$z$ and high-$z$ BAO data, and the overlapping region lies around a low $H_0$ region.  Bounds on $H_0$ and $\Omega_m$ in this framework are given as 
\begin{equation}
\label{eq:OkLCDM_H0_Om}
H_0 = 67.52 ^{+3.21}_{-3.05} \,,
\qquad 
\Omega_m = 0.293 ^{+0.029}_{-0.028} 
\qquad
[\Omega_K \Lambda{\rm CDM}] \,.
\end{equation}
Notice that because of the degeneracies of $\Omega_K$--$\Omega_m$ and $\Omega_K$--$H_0$, the uncertainty for $H_0$ is larger even when compared to that for $w$CDM and $N_{\rm eff} \Lambda$CDM models.  Therefore the significance of the $H_0$ tension in this model is 1.7$\sigma$, which is weaker than the other models.

\begin{figure}
\centering
\scalebox{0.6}{\includegraphics{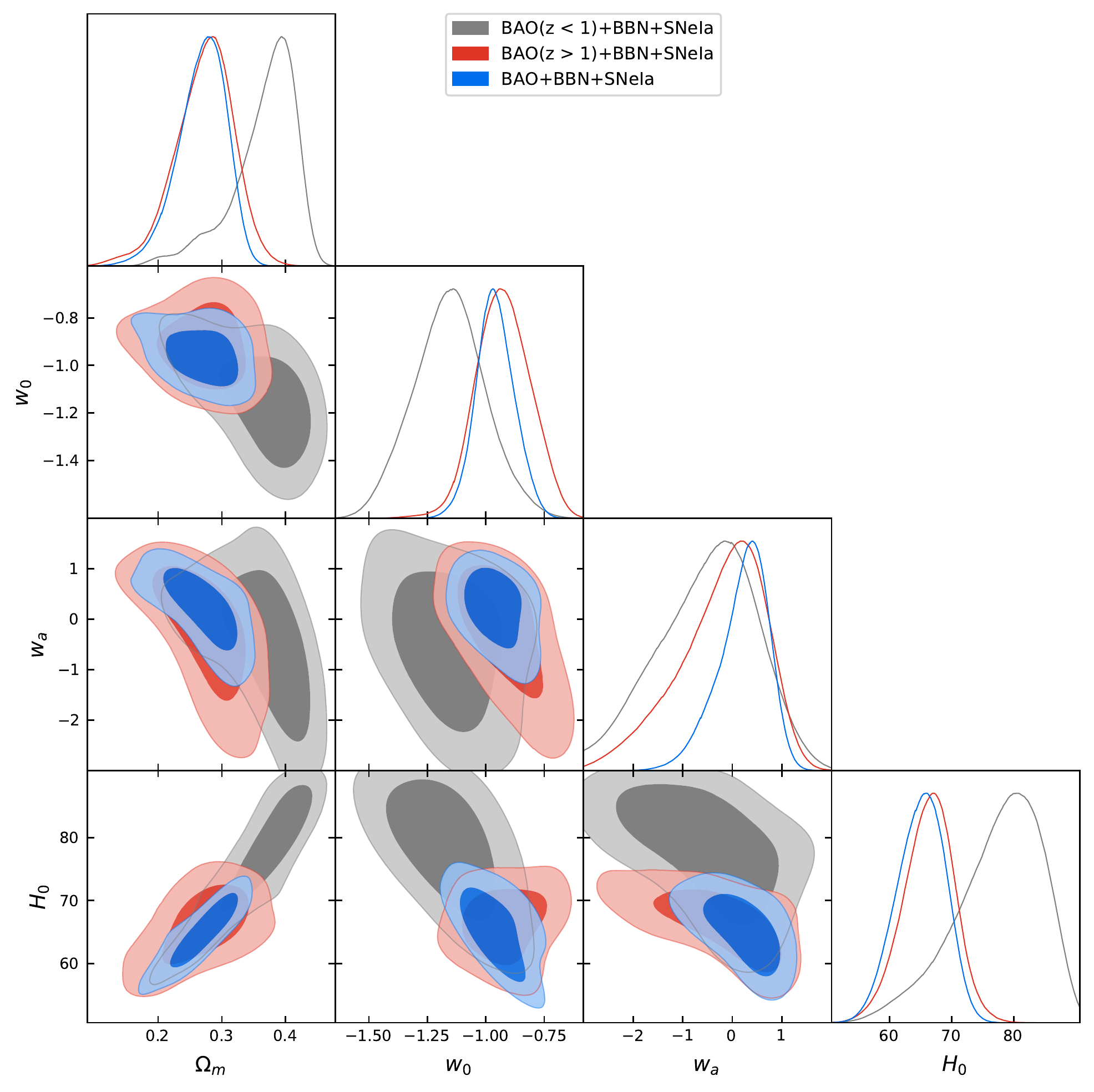}}
\caption{Constraints for $w_0 w_a$(CPL)CDM model  \label{fig:w0waCDM} }
\end{figure}

\bigskip\bigskip
\noindent {\bf $\bm{w_0 w_a}$(CPL)CDM model}: 
In Fig.~\ref{fig:w0waCDM}, the constraints in the $w_0 w_a$(CPL)CDM model where the CPL parametrization for dark energy EoS is adopted are shown. In this model, the function $f(z)$ is given by 
\begin{equation}
\left. f  (z) \right|_{w_0 w_a{\rm CDM}} = \frac{1}{\sqrt{\Omega_m (1+z)^3 + (1 - \Omega_m)(1+z)^{3(1+w_0 + w_a)} \exp \left[-\displaystyle\frac{3 w_a z}{1+z} \right] }} \,.
\end{equation}
Expanding $ \left. f  (z) \right|_{w_0 w_a{\rm CDM}}$ at low-$z$ redshifts, one obtains
\begin{eqnarray}
\left. f  (z) \right|_{w_0 w_a{\rm CDM}} &=& 
1   + \frac32 \left( -1 - w_0 + w_0 \Omega_m \right) z + {\cal O}(z^2) \,.
\end{eqnarray}
Actually $w_a$ dependence appears only from the second order in $z$ and hence $w_a$ cannot be well constrained from the low-$z$ BAO data. This is the reason why the constraint on $w_a$ from low-$z$ BAO  is very weak as can be read off from Fig.~\ref{fig:w0waCDM}. Regarding the constraints on $\Omega_m, w_0$ and $H_0$, the tendencies are the same as the ones for $w$CDM model, however, due to the degeneracy between $w_0$ and $w_a$, which also propagates to $\Omega_m$ and $H_0$, the uncertainties become larger compared to those for the $w$CDM case. The mean values and 1$\sigma$ uncertainties for $H_0$ and $\Omega_m$ in this model are 
\begin{equation}
\label{eq:CPL_H0_Om}
H_0 = 65.09 ^{+3.88}_{-3.73} \,,
\qquad 
\Omega_m = 0.269 ^{+0.038}_{-0.036} 
\qquad
[w_0 w_a{\rm (CPL)CDM}] \,.
\end{equation}
Although the mean value for $H_0$ gets  smaller than that for the $\Lambda$CDM case, the uncertainty gets larger, which makes the tension  weaker and its significance is 2.0$\sigma$.

\begin{figure}
\centering
\scalebox{0.65}{\includegraphics{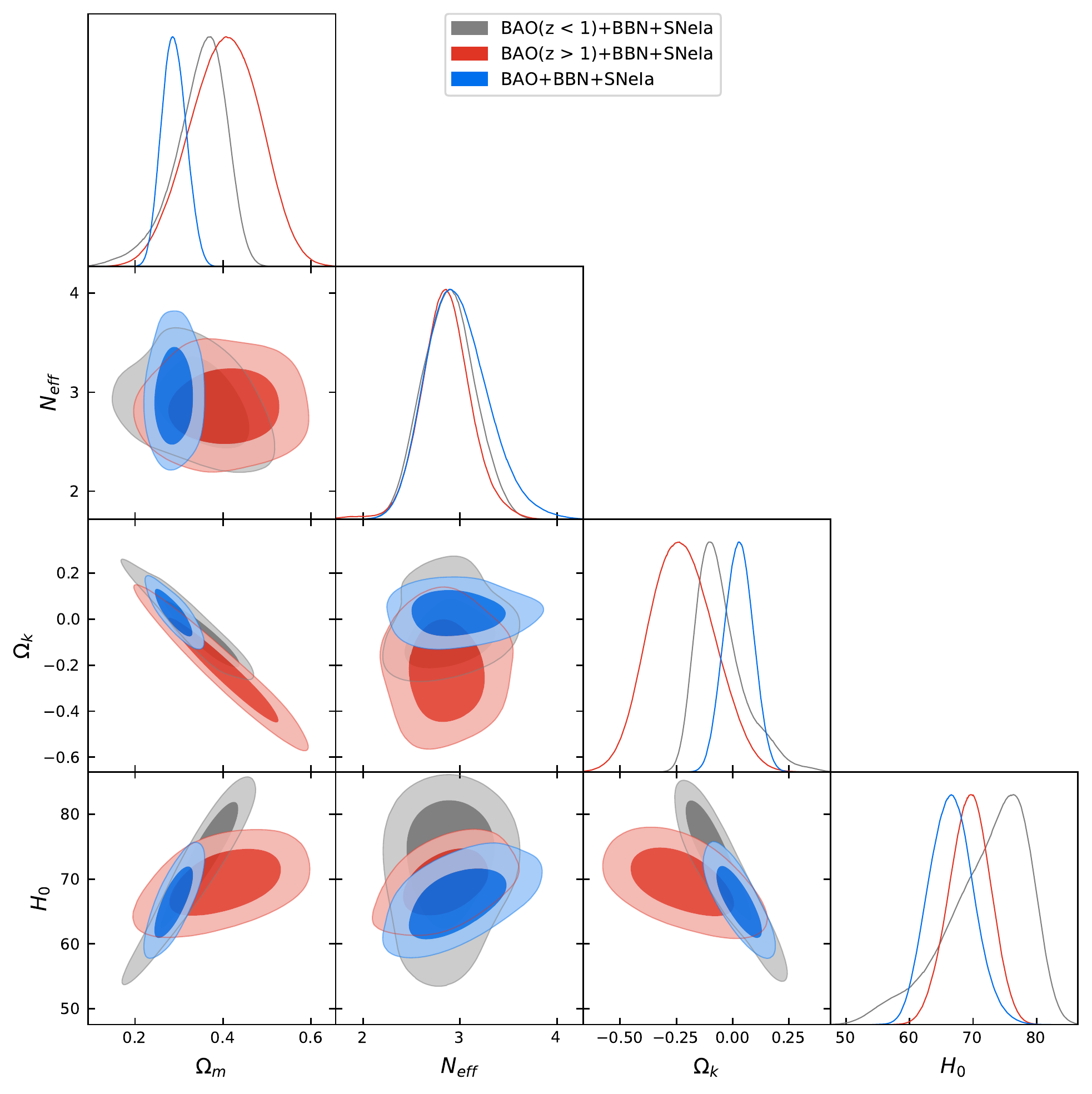}}
\caption{ Constraints for $N_{\rm eff} \Omega_K \Lambda$CDM model \label{fig:Neff_Ok_CDM} }
\end{figure}

\bigskip\bigskip
\noindent 
{\bf $\bm{N_{\rm eff} \Omega_K \Lambda}$CDM model}:  In this model, $N_{\rm eff}$ and $\Omega_K$ are allowed to vary and hence the constraints obtained in this model share the same tendency of the $N_{\rm eff} \Lambda$CDM and $\Omega_K \Lambda$CDM models discussed above as shown in Fig.~\ref{fig:Neff_Ok_CDM}.  Constraints on $H_0$ and $\Omega_m$ in this model are given as
\begin{equation}
\label{eq:NeffOkLCDM_H0_Om}
H_0 = 66.54 ^{+3.42}_{-3.39} \,,
\qquad 
\Omega_m = 0.288 ^{+0.027}_{-0.026} 
\qquad
[N_{\rm eff} \Omega_K \Lambda{\rm CDM}] \,.
\end{equation}
Although the mean value of $H_0$ gets lower compared to that for the $\Lambda$CDM model, $N_{\rm eff}$ and $\Omega_K$ are degenerate with $H_0$, and hence the uncertainty for $H_0$ becomes large  and the $H_0$ tension in this model is as modest as 1.8$\sigma$.

\begin{figure}
\centering
\scalebox{0.65}{\includegraphics{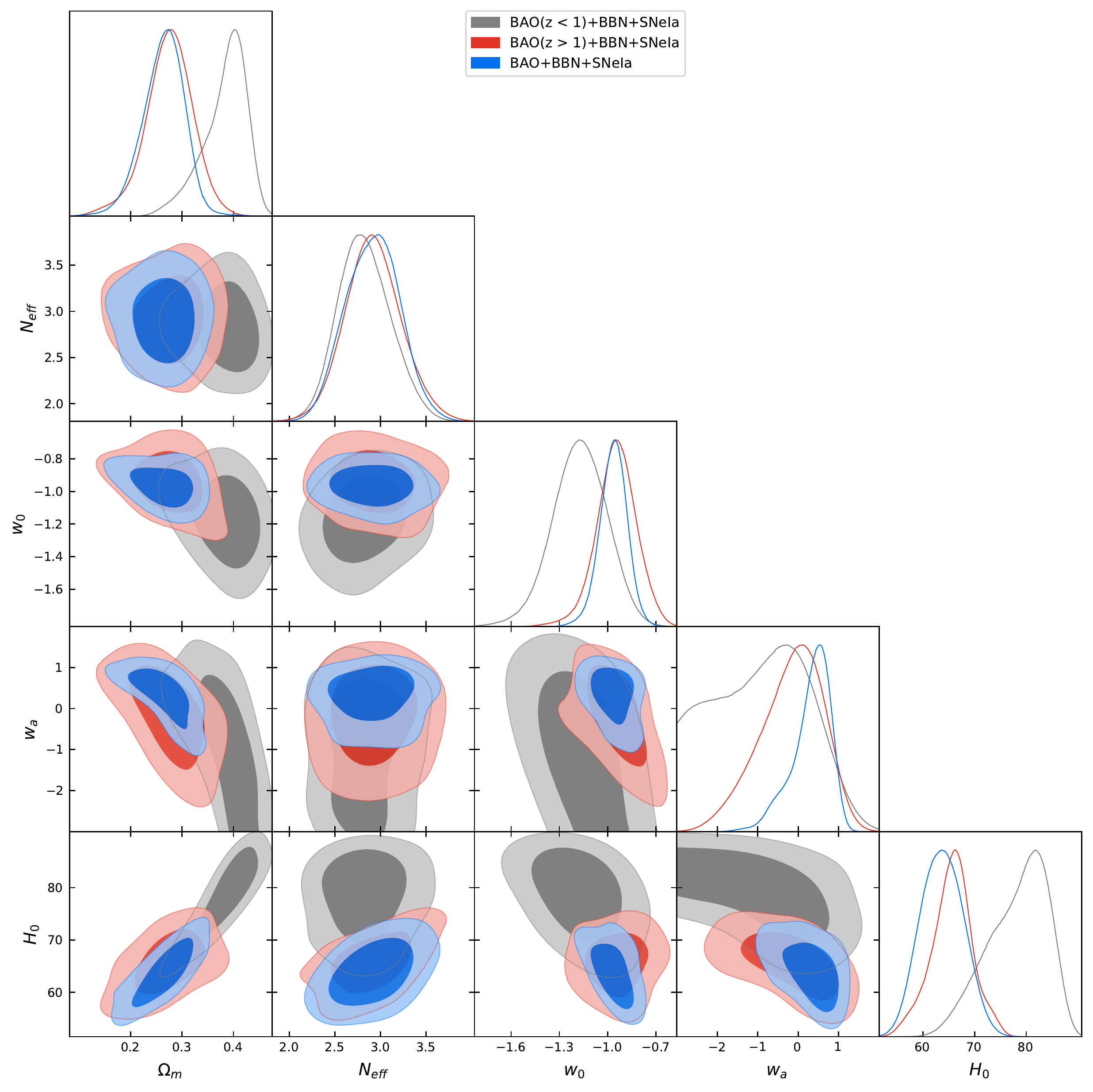}}
\caption{ Constraints for $N_{\rm eff} w_0 w_a$(CPL)CDM model \label{fig:Neff_w0waCDM} }
\end{figure}

\bigskip\bigskip
\noindent  
{\bf $\bm{N_{\rm eff} w_0 w_a} $(CPL)CDM model}:  We show the constraints for the $N_{\rm eff} w_0 w_a$CDM model in Fig.~\ref{fig:Neff_w0waCDM}.  This model shares the properties of the $N_{\rm eff} \Lambda$CDM and $w_0 w_a$CDM models discussed above.  The values of $H_0$ and $\Omega_m$ derived in this framework are 
\begin{equation}
\label{eq:NeffCPL_H0_Om}
H_0 = 63.72 ^{+4.00}_{-4.27} \,,
\qquad 
\Omega_m = 0.264 ^{+0.037}_{-0.042} 
\qquad
[N_{\rm eff} w_0 w_a {\rm (CPL) CDM}] \,.
\end{equation}
As can be noticed from Eqs.~\eqref{eq:NeffCDM_H0_Om} and \eqref{eq:CPL_H0_Om}, the mean value of $H_0$ gets lower in the $N_{\rm eff} \Lambda$CDM and $w_0 w_a$CDM models, and the same is true for this model. Actually, the $H_0$ mean value gets much lower in this framework.  Since the uncertainty for $H_0$ is already large in the $N_{\rm eff} \Lambda$CDM and $w_0 w_a$CDM models, it gets even larger in this model compared to that for the $N_{\rm eff} \Lambda$CDM and $w_0 w_a$CDM models.  However, even with such a large uncertainty, the value of $H_0$ obtained from BAO+BBN+SN is inconsistent with the local measurements at the 2.2$\sigma$ level due to the fact that the mean value becomes low in this model as described above. Therefore the $H_0$ tension still persists even in the $N_{\rm eff} w_0 w_a$CDM model although the tension is relaxed to some extent because of the large uncertainty of $H_0$ in this model.

\begin{figure}
\centering
\scalebox{0.65}{\includegraphics{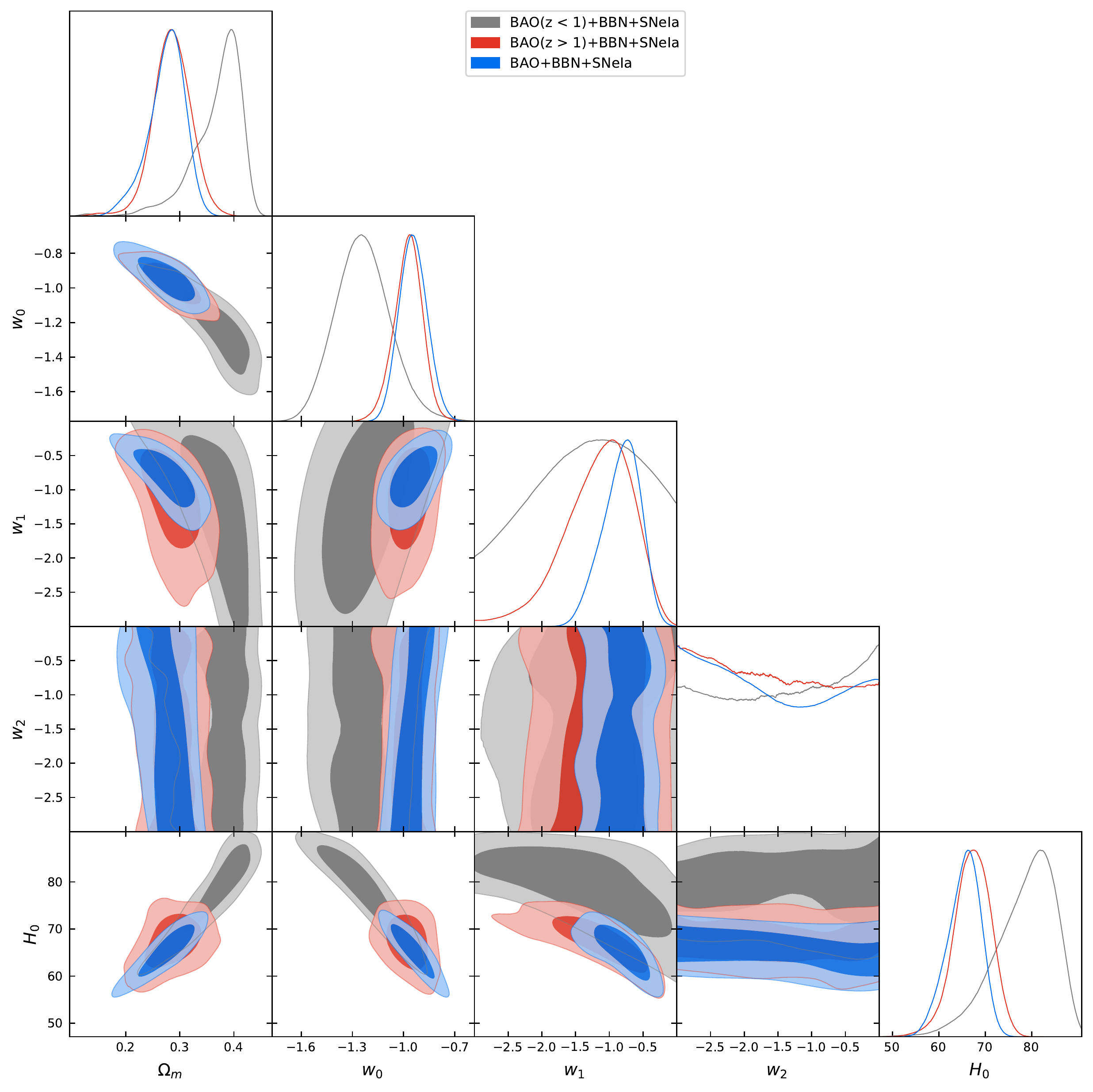}}
\caption{Constraints for $w_0 w_1 w_2$CDM model \label{fig:w0w1w2CDM}. }
\end{figure}

\bigskip\bigskip
\noindent 
{\bf $\bm{w_0 w_1 w_2}$CDM model}:  In this framework,  the EoS  for dark energy is assumed to be $w_0, w_1$ and $w_2$ for the redshift ranges of $0 < z <0.5, \,\, 0.5 < z <2$ and $ 2 <z$, respectively. Although this model includes more parameters for dark energy EoS than those for $w$CDM, constraints on $H_0$ and $\Omega_m$ share the same tendency with those as the $w$CDM model. One can see the  resemblance of the constraints in the plane of $\Omega_m$--$H_0$,  $w_0$--$H_0$ and $\Omega_m$--$w_0$ in Fig.~\ref{fig:w0w1w2CDM} with those shown in Fig.~\ref{fig:wCDM} for $w$CDM model. Since there are some degeneracies among $w_0, w_1, \Omega_m$ and $H_0$, the uncertainties for $H_0$ and $\Omega_m$ in the $w_0 w_1 w_2$CDM model become larger than those for the $w$CDM case. Since observations of SN and BAO mainly probe the evolution of the Universe at low redshift  $\left( z \lesssim 2 \right)$, $w_2$ in particular (the dark energy equation of state for $z>2$) does not affect the fit to these observations and hence $w_2$ cannot be well constrained as seen from Fig.~\ref{fig:w0w1w2CDM}. One can also notice there is almost no correlation between $w_2$ and other parameters, which can be read off from the 2D constraints in the $w_2$--$\Omega_m$,  $w_2$--$w_0$,  $w_2$--$w_1$ and $w_2$--$H_0$ planes in Fig.~\ref{fig:w0w1w2CDM}. Therefore even if $w_2$ cannot be well determined, constraints for other  parameters are not affected much. The bounds on $H_0$ and $\Omega_m$ in this model are given as 
\begin{equation}
\label{eq:DEij_H0_Om}
H_0 = 65.45 ^{+3.56}_{-3.82} \,,
\qquad 
\Omega_m = 0.276 ^{+0.033}_{-0.033} 
\qquad
[w_0 w_1 w_2{\rm CDM}] \,.
\end{equation}
Although the uncertainty for $H_0$ is relatively large in this model, the $H_0$ tension with the local measurement still persists at the $2.0 \sigma$ level.


\bigskip\bigskip
In Fig.~\ref{fig:summary}, we summarize the constraints on $H_0$ in $\Lambda$CDM and the extended models studied in this paper. For comparison, we also show the value of $H_0$ obtained from the local measurements, in which we adopt the value  from SH0ES \cite{Riess:2020fzl} [its actual value is quoted in Eq.~\eqref{eq:local_H0}].  The inner dark and outer light vertical red bands correspond to 1$\sigma$ and 2$\sigma$ bounds from the direct local measurements.  As summarized in the figure, the $H_0$ tension still persists even without CMB data in every model investigated in this paper, although the significance of the tension in the extended models is weakened compared to that in a flat $\Lambda$CDM model and varies, depending on the models, at the level of $\sim 2\sigma$.   Given that we only include the data from BAO, BBN and SN, the significance is not very large  when compared to the analysis including CMB. In any case, we can conclude that the $H_0$ tension does not go away even if we consider a range of extensions of the $\Lambda$CDM model even without CMB data.

\pagebreak
\section{Conclusion  \label{sec:conclusion}}

We investigated whether the $H_0$ tension persists (or not) even without CMB data, specifically the data from BAO, BBN and SN,  in a range of extended model frameworks beyond $\Lambda$CDM. As summarized in Fig.~\ref{fig:summary}, in every model we studied in this paper, there is still a tension between the value obtained from indirect observations used in our analysis (i.e., from BAO, BBN and SN) and the one from the local direct measurements. 

Models studied in this paper include not  only  a simple (one parameter) extension from $\Lambda$CDM like $w$CDM, $N_{\rm eff} \Lambda$CDM and $\Omega_K\Lambda$CDM models, but also a bit more complicated (but not so exotic) ones such as  $w_0 w_a {\rm (CPL)}$CDM, $N_{\rm eff} \Omega_K$CDM models, which are two parameter extensions, and $N_{\rm eff}w_0w_a$(CPL)CDM and $w_0w_1w_2$CDM ones, which are three parameter extensions. We found that the values  of $H_0$ obtained from non-CMB data in those frameworks are overall lower than that measured  in the direct observations. Although, depending on the model, the significance of the tension varies,  and in some cases, the tension is less prominent compared to the one in the $\Lambda$CDM model,  we can conclude that the $H_0$ tension exists in a broad class of models even without CMB data.

We should also mention a possibility that some systematic effects might  be the origin of the $H_0$ tension (see e.g., \cite{Efstathiou:2020wxn,Birrer:2020tax}). Such kinds of systematics might fully or partially explain the tension, however, our analysis in this paper would indicate that a low value of $H_0$ obtained by indirect measurements seem to be robust regardless of the data set adopted and the cosmological framework assumed in this paper. Therefore it would be worth further pursuing  a novel new physics in the light of $H_0$ tension, which may bring us a more complete understanding about the evolution of our Universe.

\section*{Acknowledgements}
This work is supported by JSPS KAKENHI Grants No.~18H04339~(TS), No.~18K03640~(TS), No.~17H01131~(TT, TS), 19K03874~(TT), and
MEXT KAKENHI Grant No.~19H05110~(TT).

\bibliography{ref_H0_BAO_BBN}

\end{document}